\documentclass[aps,prc,preprint,showpacs,superscriptaddress,floatfix,raggedbottom]{revtex4-1}
\usepackage{amsmath, amsthm, amssymb}
\usepackage{graphics}
\usepackage{graphicx}
\usepackage{epsfig}
\usepackage{gensymb}        % provides macro \degree which works in text and math
\usepackage{float}
\usepackage{multirow}  

\usepackage{amssymb}  
\usepackage{subfig}    
\usepackage{array}
\begin{document}
\title{Investigation of the one-neutron transfer in $^{13}$C + $^{28}$Si at E$_{\textnormal{lab}}$~=~30 and 34  MeV}

\author{R.~Linares}
\email{rlinares@id.uff.br}
\author{C.~C.~Seabra}
\author{V.~A.~B.~Zagatto}
\affiliation{Instituto de F\'isica, Universidade Federal Fluminense, 24210-340, Niter\'oi, Rio de Janeiro, Brazil}

\author{V.~Scarduelli}
\affiliation{Instituto de F\'isica, Universidade Federal Fluminense, 24210-340, Niter\'oi, Rio de Janeiro, Brazil}
\affiliation{Instituto de F\'isica, Universidade de S\~ao Paulo, S\~ao Paulo, Brazil}

\author{L.~Gasques}
\author{L.~C.~Chamon}
\affiliation{Instituto de F\'isica, Universidade de S\~ao Paulo, S\~ao Paulo, Brazil}

\author{B.~R.~Gon\c{c}alves}
\author{D.~R.~Mendes Junior}
\affiliation{Instituto de F\'isica, Universidade Federal Fluminense, 24210-340, Niter\'oi, Rio de Janeiro, Brazil}
\author{A.~L\'epine-Szily}
\affiliation{Instituto de F\'isica, Universidade de S\~ao Paulo, S\~ao Paulo, Brazil}

\date{\today}
    
\begin{abstract} 

\textbf{Background:} Neutron transfer measurements for the $^{18}$O + $^{28}$Si system have shown that the experimental one-neutron and two-neutron transfer cross sections are well reproduced with spectroscopic amplitudes from two different shell model interactions for the Si isotopes: \textit{psdmod} for the two-neutron transfer, and \textit{psdmwkpn} for the one-neutron transfer. 
%As its origin may be related to the pre-formed paired neutrons of projectile, studying a nucleus where this characteristic is absent (like $^{13}$C) should help to ellucidate this question.
%The origin of this ambiguity was not clear but can be related with the pre-formed paired neutrons in the $^{18}$O projectile that requires a description of pairing effect in the one neutron ($^{18}$O,$^{17}$O) transfer reaction. Such paired neutrons are not present in projectiles such as $^{13}$C. 

\textbf{Purpose:} The origin of this ambiguity can be related to a more complex mechanism in the one-neutron transfer that requires the unpairing of neutrons prior to its transfer in the ($^{18}$O,$^{17}$O) reaction. Studying a nucleus where this characteristic is absent ($^{13}$C) should help to elucidate this question.
%The one-neutron spectroscopic factors of the Si isotopes are investigated using the ($^{13}$C,$^{12}$C) reaction.

\textbf{Method:} One-neutron transfer cross sections were measured for the $^{13}$C + $^{28}$Si at E$_{\textnormal{lab}}$ = 30, and 34 MeV, and compared with coupled reaction channel calculations using spectroscopic amplitudes derived from the \textit{psdmod} and \textit{psdmwkpn} shell model interactions. 

\textbf{Results:} The spectroscopic amplitudes from the \textit{psdmod} interaction for the relevant states in $^{29}$Si provide a good description of the experimental data and the corresponding values agree with previous estimates obtained from the (d,p) reaction.

\textbf{Conclusions:} The experimental data for the one-neutron transfer to $^{28}$Si induced by ($^{13}$C,$^{12}$C) reaction is well reproduced using spectroscopic amplitudes from the \textit{psdmod}.

\end{abstract}


\maketitle

\section{Introduction}
\label{Intro}

Particle configurations of bound states in the atomic nuclei can be studied using transfer reactions. In such studies, the optical potential and spectroscopic factors (${\cal{S}}^2$) are important parameters in the calculations of the transfer cross sections. Experimental values for ${\cal{S}}^2$ can be obtained from a direct comparison between experimental and theoretical cross sections, as in (d,p) \cite{MWC71, PFR83}, (t,d) \cite{PCG87}, ($^7$Li,$^6$Li) \cite{SUD73} reactions. However, the experimental approach may lead to some ambiguities in the ${\cal{S}}^2$ values for many nuclei. For instance, the ${\cal{S}}^2$ value for a p$_{1/2}$ valence neutron to $^{12}$C ranges from 0.3 (from $^{13}$C(p,d)$^{12}$C data at 65 MeV \cite{HKS80}) to 1.4 (from $^{12}$C(d,p)$^{13}$C data at 15 MeV \cite{DSH73}). The main reasons for these discrepancies are the adopted optical potentials and the coupling scheme considered in direct reaction calculation \cite{LFL04}.

Recent advances in experimental setups have renewed the use of heavy-ion probes, like ($^{13}$C,$^{12}$C) and ($^{18}$O,$^{17}$O), in transfer reactions \cite{CAC18}. Some advantages over the use of light-ion are: i) avoid the inclusion of the break-up channel \cite{BTT16}; and ii) suppression of the effect of non-locality, that is relevant in (d,p) reactions \cite{TiJ13, RTN16}. On the other hand, effects of strong absorption are more pronounced and the angular distributions exhibit a diffraction-like pattern as the bombarding energy increases. Moreover, second-order mechanisms such as projectile/target excitation preceding and/or following the transfer of nucleons, must be taken into account properly. In addition, partial waves that contribute to the transfer reaction are limited to a range of optimum Q values for a given reaction and energy.

%dealing with heavy-ion systems requires a complete coupling scheme that also includes second-order transfer, such as target excitation preceding or following the transfer of nucleon.

Recently, analysis of the one-neutron (1NT) and two-neutron transfer (2NT) to $^{28}$Si, induced by the ($^{18}$O,$^{17}$O) \cite{LEL18} and ($^{18}$O,$^{16}$O) \cite{CLL18} reactions respectively, have been reported. In these works, coupled reaction channels (CRC) were performed including  ${\cal{S}}^2$ for the relevant states derived from nuclear shell models with suitable interactions and model spaces to describe the low-lying states in the $^{28,29,30}$Si isotopes. Best agreement between experimental data and calculations have been achieved adopting different interactions for the 1NT (\textit{psdmwkpn}) and 2NT (\textit{psdmod}) processes. 

%There are some questions that shall be addressed:

%\begin{itemize}
%    \item Does the pre-formed paired valence neutrons in the $^{18}$O nucleus affect the 1NT probabilities?
%    \item Does the static deformation of the target nucleus affects the 1NT mechanism?
%\end{itemize}

It is not clear how the 1NT is affected by the pre-formed paired valence neutrons in the $^{18}$O nucleus. In this work, we have analyzed the 1NT to the $^{28}$Si using the ($^{13}$C,$^{12}$C) probe. The $^{13}$C is treated as a single valence neutron bound to a $^{12}$C nuclear core. We have measured elastic cross sections at E$_{\textnormal{lab}}$ =~25, 30, and 34 MeV and inelastic and one-neutron transfer cross sections at E$_{\textnormal{lab}}$ =~30, and 34 MeV. Cross sections for elastic and inelastic scattering are used to constraint the parameters of the effective nucleus-nucleus potential. This paper is organized as follows: the experimental details and the theoretical analysis are discussed in sections~\ref{exp} and~\ref{theor}, respectively. The conclusions are given in section~\ref{conc}.

%In this system, the Q-value for one-neutron transfer is +3.53 MeV and, therefore, can be energetically distinguished from elastic and inelastic events. 

% = = = = = EXPERIMENTAL SECTION = = = = = = 
\section{\label{exp}Experimental details}

The experiment was performed at the 8 MV tandem accelerator of the University of S\~ao Paulo. The $^{13}$C beam was accelerated at E$_{\textnormal{\textnormal{lab}}}$~=~25, 30, and 34 MeV with averaged beam intensity of about 30 enA on the target. For the $^{13}$C + $^{28}$Si system, the Coulomb barrier height is V$_{\textnormal{B}}$~=~18.9~MeV (in the laboratory reference).

%%%%%%%%%%%%%%%%%%%%%%%%%%%%%%%%%%%%
\begin{figure}[tb!]
\centering
\graphicspath{{figuras/}}
\includegraphics[width=0.45\textwidth]{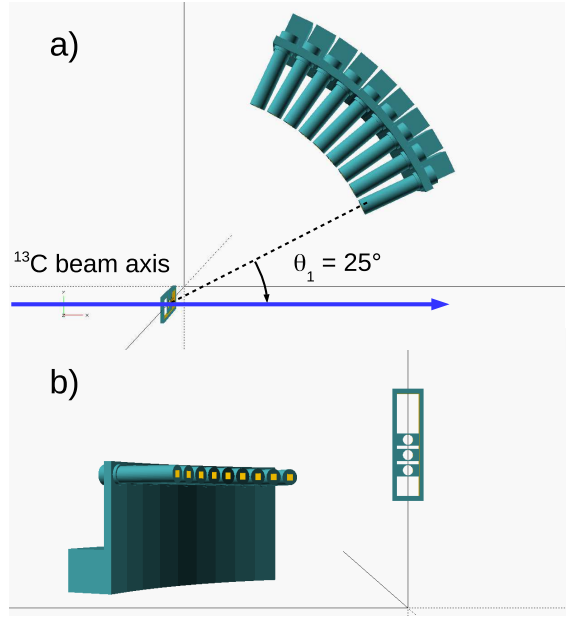}
\caption{(Color online) Sketch of the experimental setup with an array of 9 Si detectors mounted in angular steps of $5\degree$ with the first one at $\theta_{\textnormal{lab}} = 25\degree$. In panel a) top view of the setup and b) frontal view.} 
\label{Fig:setup}
\end{figure}
%%%%%%%%%%%%%%%%%%%%%%%%%%%%%%%%%%%%

Fig.~\ref{Fig:setup} shows a sketch of the Silicon Array and Telescopes of Usp for Reactions and Nuclear applications (SATURN) system \cite{GFC18}, mounted in the scattering chamber for the measurements. A set of 9 surface barrier Si detectors was mounted at 30~cm away from the targets and with $5\degree$ of angular step size, covering the angles from $25\degree$ to $65\degree$ (laboratory framework). A 4-position mobile target ladder, placed at the center of the chamber, was mounted with 2 thin $^{28}$Si foils (Si-only) 99.9\% isotopically enriched and 2 other foils composed of $^{28}$Si with a thin backing layer of $^{197}$Au (Si+Au). Thicknesses of the $^{28}$Si layers were measured by Rutherford backscattering (RBS) of $^4$He beam and were approximately 40~$\mu$g/cm$^2$.

At each beam energy, measurements were carried out with the Si+Au target, for normalization of the cross sections, and the Si-only target, for a clear identification of the 1NT. The energy calibration of each Si detector was performed adopting the elastic peaks associated to the $^{13}$C scattered off the $^{28}$Si and $^{197}$Au nuclei. The energy resolution achieved was 0.2~MeV. The ratio between Au and Si foil thicknesses was determined from measurements at $E_{\textnormal{lab}} = 25$~MeV, using the angular points at $\theta_{\textnormal{lab}}$ = 25$\degree$, 30$\degree$ and 35$\degree$, and the theoretical curve. 

\begin{figure}[tb]
\centering
\graphicspath{{figuras/}}
\includegraphics[width=0.45\textwidth]{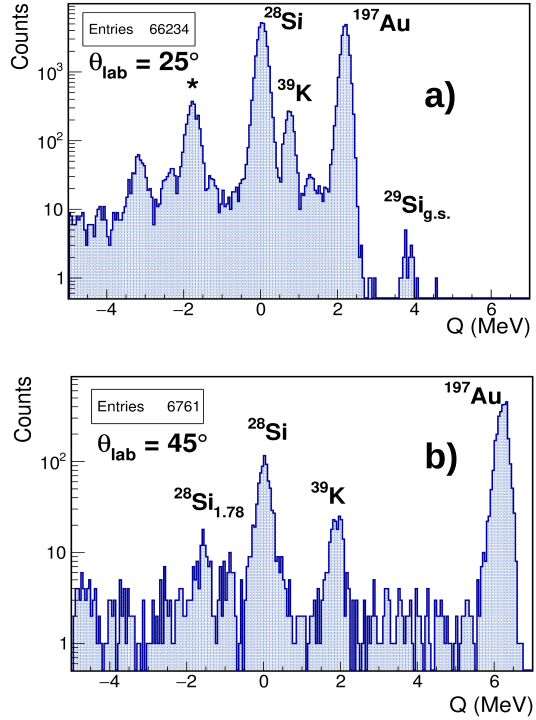}
\caption{(Color online) The $Q$-energy spectrum for the measurements with $^{13}$C at 34~MeV on the Si+Au target. The spectra observed at $\theta_{\textnormal{\textnormal{lab}}}$ = $25\degree$ and $45\degree$ are shown in panel a) and b), respectively. The asterisk in panel a) indicates the presence of oxygen contamination (in the target) in the inelastic peak.} 
\label{Fig:Q_spectrum}
\end{figure}

Typical $Q$-energy spectra, defined as the energy relative to the elastic scattering in the $^{13}$C + $^{28}$Si, are shown in Fig.~\ref{Fig:Q_spectrum} at a) $\theta_{\textnormal{lab}}$ = $25\degree$, and b) $45\degree$ measured at E$_{\textnormal{lab}}$ = 34 MeV. In this representation, the elastic scattering from $^{28}$Si corresponds to a peak centered at $Q$ = 0.0~MeV. The inelastic peak associated to the excitation of $^{28}$Si$_{1.78}$~MeV corresponds to the peak around $Q$ = -1.8~MeV. At forwards angles, scattering off contaminants (oxygen) present in the target superimposes with the inelastic peak, as indicated by the asterisk in Fig.~\ref{Fig:Q_spectrum}a. Scattering off the $^{197}$Au corresponds to peaks with $Q > 0$ with $Q$-energy that depends on the scattering angle.

For the $^{13}$C + $^{28}$Si system, the \textit{Q-value} for 1NT g.s.~$\rightarrow$~g.s. is +3.53 MeV and, therefore, this reaction channel is energetically distinguished from elastic and inelastic events. In Fig.~\ref{Fig:Q_spectrum}a, the peak at $Q$ = +3.6 MeV ($^{29}$Si$_{g.s.}$) is associated to the 1NT g.s.~$\rightarrow$~g.s.. At $\theta_{\textnormal{lab}}$ = $45\degree$, the inelastic excitations of the $^{197}$Au interfere with the $^{29}$Si$_{g.s.}$ peak (see Fig.~\ref{Fig:Q_spectrum}b). Between the elastic of $^{28}$Si and $^{197}$Au it is also observed a peak that comes from $^{39}$K contamination in the target carried in during the manufacturing process of the thin films. The presence of this contamination was also confirmed with the Rutherford Backscattering Spectrometry (RBS) analysis of the foils produced in the same batch, indicating a $2\%$ K contamination in the target.

%%%%%%%%%%%%%%%%%%%%%%%%%%%%%%%%%%%%
\begin{figure}[!tb]
\centering
\graphicspath{{figuras/}}
\includegraphics[width=0.45\textwidth]{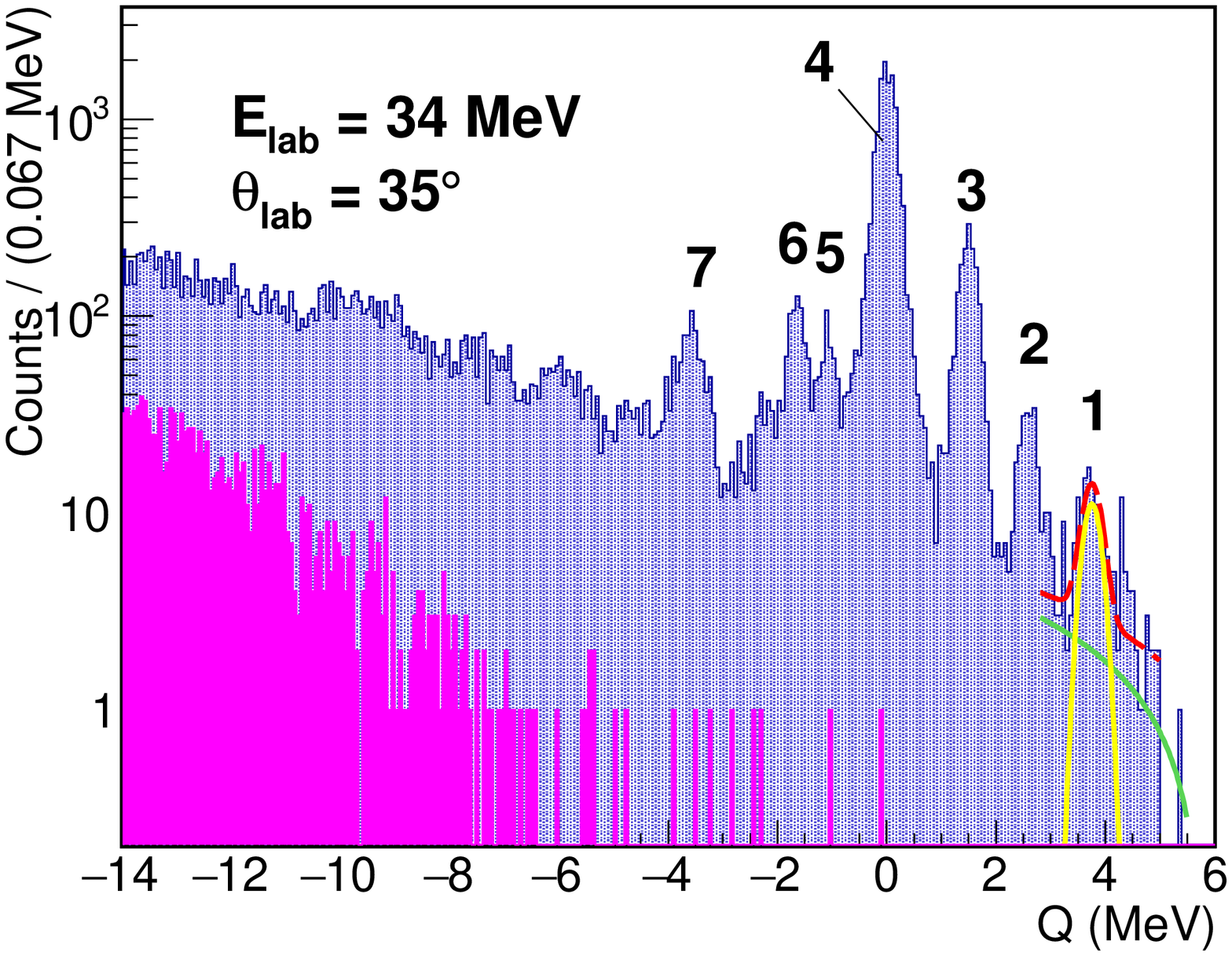}
\caption{(Color online) The $Q$-energy spectrum for the measurements with $^{13}$C at 34~MeV on the Si-only target at $\theta_{\textnormal{lab}}$ = $35\degree$ (blue histogram). In the same plot is shown the energy spectrum of p and $\alpha$ from fusion-evaporation process (purple histogram). Peak numbering as follows: 1 - $^{29}$Si$_\textnormal{g.s.}$ (1NT); 2 - $^{29}$Si$_\textnormal{1.27}$ and $^{40}$K$_{\textnormal{g.s.}}$ (1NT); 3 - elastic peak $^{39}$K; 4 - elastic peak $^{28}$Si; 5 - possibly inelastic peak in $^{39}$K$_{2.52}$; 6 - inelastic peak $^{28}$Si$_{1.78}$; 7 - elastic peak $^{16}$O and $^{12}$C. Counts in the peak 1 were determined from the gaussian curve represented by the yellow curve. See text for further details. }
\label{Fig:Q_spectrum_Si}
\end{figure}
%%%%%%%%%%%%%%%%%%%%%%%%%%%%%%%%%%%%

A typical spectrum for Si-only target is shown in Fig.~\ref{Fig:Q_spectrum_Si} (blue histogram). The 1NT to the $^{29}$Si$_{\textnormal{g.s.}}$, the elastic peak of the $^{28}$Si and inelastic peak of the $^{28}$Si$_{1.78}$ correspond to peaks 1, 4 and 6, respectively. Other peaks associated to reactions with contaminants on the target were also identified (listed in the caption of Fig.~\ref{Fig:Q_spectrum_Si}), except the peak 5. This is possibly associated to the inelastic scattering that populates the 1/2$^+$ (2.52~MeV) in the $^{39}$K. Calculation of the energies of p and $\alpha$ particles coming from the fusion-evaporation process, using the PACE4 code \cite{TaB08,Gav80}, is presented in Fig.~\ref{Fig:Q_spectrum_Si} (purple histogram). This shows that, in the energy range of the $^{29}$Si$_{g.s.}$ peak, there is no significant interference of high energetic p or $\alpha$ particles.

Yields in the elastic, inelastic and 1NT were determined from a gaussian curve on top of a linear background fitted to the experimental peaks. An example of such fits is shown in Fig.~\ref{Fig:Q_spectrum_Si} in which the gaussian (yellow curve) and the linear (green curve) components are reproduced for the peak number 1. There are some counts with Q-value higher than +4.0 MeV that may come from contaminant heavier than K in the the Si-only target. Possible heavy contaminants are $^{127}$I, from the release agent, and $^{184}$W, from the cathode used as holder for the $^{28}$Si powder for the manufacturing of the thin films. Such heavier contaminants were not observed in the RBS analysis. Even though, in both cases the elastic scattering of $^{13}$C would produce peaks at Q-values +4.1 and +4.5 MeV, respectively. For the inelastic peak, the background due to the presence of peak 5 at some angles was subtracted adopting a linear behavior.

\section{\label{theor}Theoretical Analysis}

\begin{figure}[tb]
\centering
\graphicspath{{figuras/}}
\includegraphics[width=0.45\textwidth]{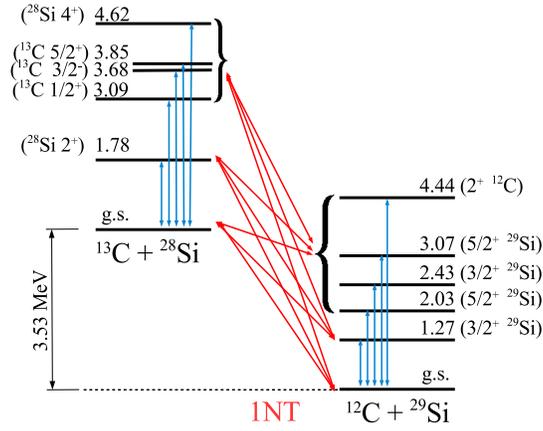}
\caption{(Color online) Coupling scheme considered in the CRC calculations.} 
\label{Fig:coupling_scheme}
\end{figure}

Direct reaction calculations were performed within the coupled reaction channel (CRC) framework  using the FRESCO code \cite{Tho88} with exact finite range and prior representation. Non-orthogonality corrections and full complex remnant terms were considered in the coupled channel equations. A sketch of the coupling scheme considered in the CRC calculations is shown in Fig.~\ref{Fig:coupling_scheme}. The inelastic channels were considered using the deformation parameter for the collective states. For the $^{28}$Si target nucleus, $\beta_2 = 0.407$, taken from Ref.~\cite{RNT01}, and for the $^{13}$C projectile nucleus, $\beta_1 = 0.143$ \cite{Ajz91}. The single-particle wave functions used in the matrix elements were generated using Woods-Saxon potential with depth adjusted to reproduce the experimental separation energies for one neutron in $^{13}$C (S$_n$ = 4.95~MeV) and $^{29}$Si (S$_n$ = 8.45~MeV). The reduced radii and diffuseness parameters for the $^{28}$Si and $^{12}$C cores were set to values previously used in the analysis with the $^{18}$O projectile. These values are 1.26 fm and 0.65 fm for $^{28}$Si core \cite{CLL18} and 1.25 fm and 0.80 fm for $^{12}$C \cite{CCB13}, respectively. Calculations have been performed within 10\% deviation in the adopted reduced radii and diffuseness values and no significant effect were observed in the results. 

%%%%%%%%%%%%%%%%%%%%%%%%%%%%%%%%%%%%
%%%%%%%%%%%%%%%%%%%%%%%%%%%%%%%%%%%%

The ${\cal {S}}^2$ values were obtained using the NuShellX code \cite{Rae08}. For $^{12,13}$C, the calculations were performed using the \textit{psdmod} interaction, that is a modified version of the \textit{psdwbt} interaction \cite{UC11}, which gives a reasonable description of the p-sd-shell nuclei.
%, which $^{4}$He core and the valence subspace for neutrons and protons in the 1p$_{1/2}$, 1p$_{3/2}$, 1d$_{3/2}$, 1d$_{5/2}$, 2s$_{1/2}$ orbitals.
For $^{28,29}$Si isotopes, two interactions are considered: again the \textit{psdmod} and the \textit{psdmwkpn} interaction \cite{BaD13}. The latter is a combination of the Cohen-Kurath interaction \cite{CoK65} for the p-shell, the Wildenthal interaction \cite{Wil84} for the sd-shell and the Millener-Kurath interaction \cite{MiK75} for the coupling matrix elements between p- and sd-shells. In both interactions, the model space assumes $^{4}$He as a closed core and valence neutrons and protons in the 1p$_{3/2}$, 1p$_{1/2}$, 1d$_{5/2}$, 1d$_{3/2}$, and 2s$_{1/2}$ orbitals. The spectroscopic amplitudes of states in $^{29}$Si for both interactions can be found in Ref.~\cite{LEL18}. For clarity, from now on CRC-\textit{psdmod} and CRC-\textit{psdmwkpn} stand for the CRC calculations using the ${\cal {S}}^2$ for the $^{29}$Si derived from \textit{psdmod} and \textit{psdmwkpn} interactions, respectively.

%%%%%%%%%%%%%%%%%%%%%%%%%%%%%%%%%%%%
\begin{figure}[tb]
\centering
\graphicspath{{figuras/}}
\includegraphics[width=0.45\textwidth]{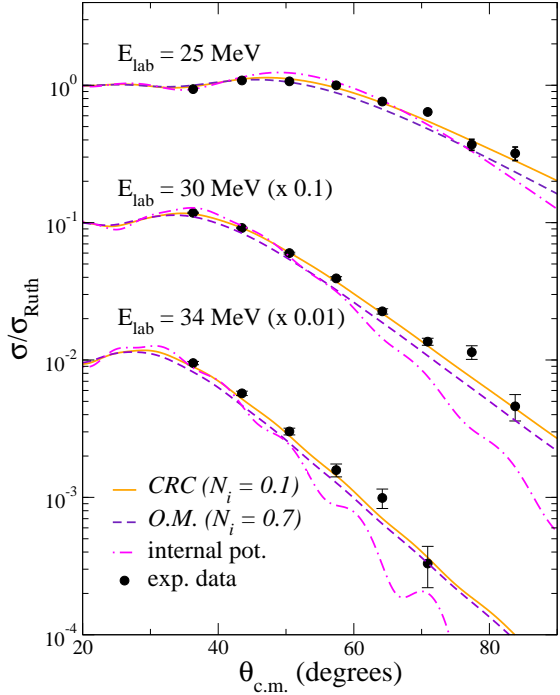}
\caption{(Color online) Angular distributions of the elastic cross sections for the $^{13}$C + $^{28}$Si at E$_{\textnormal{lab}}$ =~25, 30, and 34 MeV. The data points at $\theta_{\textnormal{lab}}$~=~25$\degree$, 30$\degree$, and 35$\degree$ for E$_{\textnormal{lab}}$~=~25~MeV were adopted for normalization of the cross sections.} 
\label{Fig:x-section_elastic}
\end{figure}
%%%%%%%%%%%%%%%%%%%%%%%%%%%%%%%%%%%%

For the CRC, the S\~ao Paulo double folding potential (SPP) \cite{CCG02} was used for the real and imaginary parts of the optical potential. In the entrance partition the N$_i$ was adjusted to describe the experimental data for elastic and inelastic scatterings to account for couplings not explicitly considered in the coupling scheme.

\newpage

Fig.~\ref{Fig:x-section_elastic} shows the angular distributions of the elastic cross sections for E$_{\textnormal{lab}}$ =~25, 30, and 34~MeV. Optical model calculations using an internal imaginary potential is shown as dot-dashed purple curve. The internal imaginary potential was defined as a Wood-Saxon shape with depth, reduced radius and diffuseness set to 50~MeV, 1.06~fm and 0.2~fm, respectively. This optical potential underestimates the cross sections at large scattering angles. A second optical model calculation was performed using the SPP shape for the imaginary part with adjustable N$_i$ factor. The best agreement between experimental data and theoretical curves is achieved for N$_i$~=~0.7, in Fig.~\ref{Fig:x-section_elastic} represented as dashed blue curves. In the CRC calculations, experimental data are well reproduced with N$_i$~=~0.1 in the entrance optical potential. Similar results are obtained for N$_i$~=~0.2, and 0.3 (not shown in Fig.~\ref{Fig:x-section_elastic}). This indicates that most relevant reaction channels (inelastic and 1NT) are accounted for in the coupling scheme and, consequently, a smaller imaginary factor is required.

\begin{figure}[tb]
\centering
\graphicspath{{figuras/}}
\includegraphics[width=0.45\textwidth]{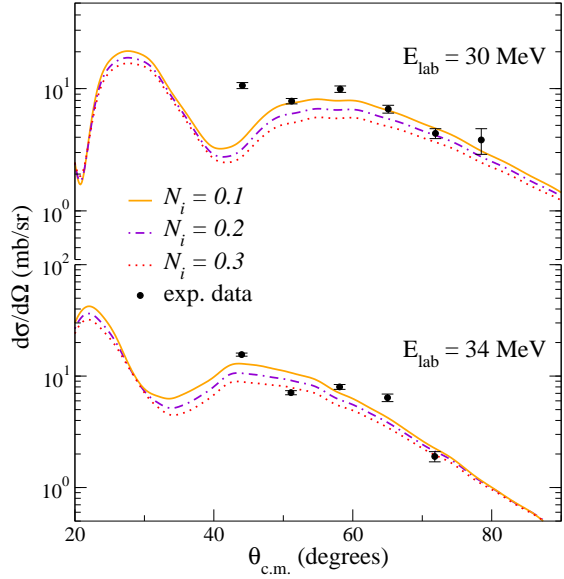}
\caption{(Color online) Angular distributions of the inelastic $2^+$ in $^{28}$Si cross sections for the $^{13}$C + $^{28}$Si at E$_{\textnormal{lab}}$ = 30, and 34  MeV.} 
\label{Fig:x-section_inelastic}
\end{figure}

The angular distributions of the inelastic cross sections to the $2^+$ excited state in $^{28}$Si for E$_{\textnormal{lab}}$ = 30, and 34~MeV are shown in Fig.~\ref{Fig:x-section_inelastic}, along with CRC calculations with different N$_i$ in the imaginary term of the optical potential. Good overall agreements between experimental data and CRC calculations are achieved for N$_i$~=~0.1 and 0.2. The theoretical curves for the elastic scattering with these N$_i$ are almost indistinguishable. The fit to elastic and inelastic data provides a good constrain to the parameter of the imaginary potential.

%Figs.~\ref{Fig:x-section_elastic} and \ref{Fig:x-section_inelastic} show the angular distributions of the elastic and inelastic cross sections to the $2^+$ excited state in $^{28}$Si, respectively, for E$_{\textnormal{lab}}$ =25, 30, and 34~MeV. along with the CRC calculations with different N$_i$ in the entrance optical potential. The inelastic cross sections are more sensitive to the factor N$_i$. Good overall agreements between experimental data and CRC calculations are achieved for N$_i$~=~0.1 and 0.2. 

%Optical model calculations using these coefficients provide a good description of the elastic scattering cross section for many systems in a wide energy interval \cite{GCG06,PLO12,OCC13}.
%The theoretical absolute cross sections for two-neutron transfer are almost insensitive to this parameter, with deviations lower than 5\%.

%%%%%%%%%%%%%%%%%%%%%%%%%%%%%%%%%%%%
%%%%%%%%%%%%%%%%%%%%%%%%%%%%%%%%%%%%

\begin{figure}[]
\centering
\graphicspath{{figuras/}}
\includegraphics[width=0.45\textwidth]{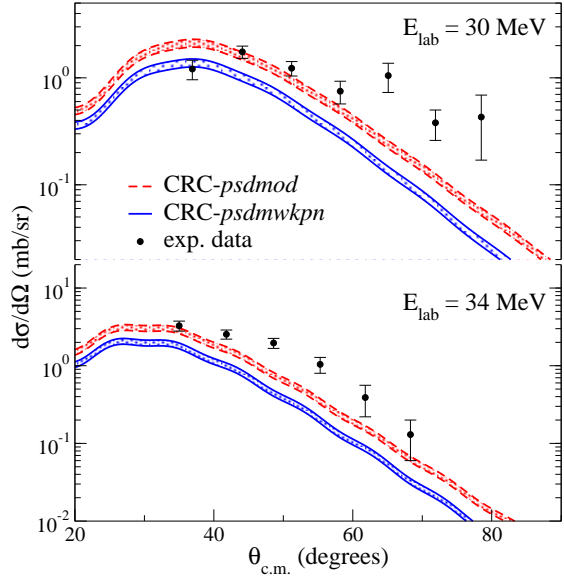}
\caption{(Color online) Angular distributions of the 1NT leading to the population of the g.s. in $^{29}$Si cross sections for the $^{13}$C + $^{28}$Si at E$_{\textnormal{lab}}$ = 30, and 34  MeV.} 
\label{Fig:x-section_1ntransfer}
\end{figure}

The cross sections for the 1NT at E$_{\textnormal{lab}}$ = 30, and 34  MeV are shown in Fig.~\ref{Fig:x-section_1ntransfer}. The CRC-\textit{psdmod} and CRC-\textit{psdmwkpn} calculations were performed using N$_i$ = 0.1 in the optical potential of the entrance partition. Similar results are obtained using N$_i$~=~0.2 and 0.3, meaning that the effect of N$_i$ values, between 0.1 and 0.3, is not strong to the 1NT channel. In the exit partition, the imaginary strength factor (N$_i$) was set 0.78, since this value provides a good description of the elastic scattering cross section for many systems in a wide energy interval \cite{ACH03}. The effect of reduced radii and diffuseness parameters, used in the form-factor to construct the single-particle wave functions of the $^{13}$C and the $^{29}$Si, has been checked. The reduced radii and diffuseness values were varied within the 1.20 - 1.25~fm and 0.7 - 0.8~fm ranges, respectively. These are represented in the envelope curves, for each CRC calculation, in Fig.~\ref{Fig:x-section_1ntransfer}. The theoretical curves are more sensitive to the diffuseness parameter. Even though, the overall effect in the calculations is not so crucial and the CRC-\textit{psdmwkpn} curves systematically lie below the CRC-\textit{psdmod}. The coupling space has also been checked and the results for elastic, inelastic and 1NT are practically the same with the removal of $3/2^-$ and $5/2^+$ states in $^{13}$C and the $4^+$ state in $^{28}$Si.

%%%%%%%%%%%%%%%%%%%%%%%%%%%%%%%%%%%%
%%%%%%%%%%%%%%%%%%%%%%%%%%%%%%%%%%%%

The CRC-\textit{psdmod} reproduces better the  experimental values at E$_{\textnormal{lab}}$ = 34 MeV and the agreement is limited at 30 MeV. Nevertheless, the CRC-\textit{psdmwkpn} underestimates the cross sections at both energies. This indicates that the \textit{psdmod} interaction provides a better estimate for the ${\cal {S}}^2$ of the $^{28,29}$Si isotopes. In Table~\ref{tab:spectroscopic}, it is presented a comparison between the spectroscopic factors (${\cal {S}}^2$) for the  $^{28}$Si to $^{29}$Si transitions derived from the \textit{psdmod} and \textit{psdmwkpn} interactions and experimental estimates obtained from the (d,p) reaction and DWBA calculations \cite{MWC71,PFR83, TLL05}. The value of ${\cal {S}}^2$ from \textit{psdmod} is close to the one reported in Ref.~\cite{MWC71} whereas the \textit{psdmwkpn} estimate is closer to that in Ref.~\cite{PFR83}. All values are within the 1-uncertainty interval obtained from a systematic analysis of experimental data for (d,p) and using a deuteron optical potential which approximately accounts for deuteron breakup Ref.~\cite{TLL05} .

%%%%%%%%%%%%%%%%%%%%%%%%%%%%%%%%%%%%%%%%%%%%%
\begin{table*} [t]
\caption{Spectroscopic factors (${\cal {S}}^2$) for the $^{28}$Si to $^{29}$Si transitions obtained by shell model calculations using \textit{psdmod} and \textit{psdmwkpn} interactions. Values obtained from $^{28}$Si(d,p)$^{29}$Si of Refs.~\cite{MWC71,PFR83, TLL05} are also included.} 
%\footnotesize
\centering
\begin{tabular}{ c c c c c c c }
\hline
 \multicolumn{2}{c}{\textbf{states}}  & \multicolumn{5}{c}{${\cal {|S|}}^2$} \\
initial & final & \textit{psdmod} & \textit{psdmwkpn} & \text{Ref.~\cite{MWC71}} & \text{Ref.~\cite{PFR83}} & \text{Ref.~\cite{TLL05}} \\ \hline
\hline

$^{28}$Si$_{\textnormal{g.s.}}$ & $^{29}$Si$_{g.s.}$  & 0.51 & 0.32 & 0.53 & 0.37 & 0.42 $\pm$ 0.13 \\ 
%        & $^{29}$Si$_{1.27}$  & 0.68 & 0.65 & 0.74 & 0.53 &  -   \\ 
%        & $^{29}$Si$_{2.03}$  & 0.12 & 0.20 & 0.12 & 0.09 &  -   \\
%        & $^{29}$Si$_{2.43}$  & 0.002&  -   &   -  &   -  &  -   \\
%        & $^{29}$Si$_{3.07}$  & 0.05 & 0.06 & 0.06 & 0.03 &  -   \\ 
\hline

\end{tabular}
\label{tab:spectroscopic}
\end{table*}
%%%%%%%%%%%%%%%%%%%%%%%%%%%%%%%%%%%%%%%%%%%%%

%Reliable theoretical predictions for transfer reactions demand explicitly inclusion of relevant reaction channels, accurate optical potential to account for channels not included in the couplings and a description of the nuclear structure of the nuclear partners in the partitions. 
The success of CRC-\textit{psdmod} compared to the present data is consistent with analysis for the 2NT in the $^{28}$Si($^{18}$O,$^{16}$O)$^{30}$Si, for which the experimental data were reproduced adopting the ${\cal {S}}^2$ derived from the \textit{psdmod} interaction for the Si isotopes \cite{CLL18}. In the analysis of the 1NT to $^{28}$Si induced by ($^{18}$O,$^{17}$O) reaction, the experimental data were reproduced better using the ${\cal {S}}^2$ from the \textit{psdmwkpn} interaction \cite{LEL18} instead. The fact that different shell model interactions are adopted for the description of the 1NT and 2NT experimental data are interpreted as follows. Accurate prediction for transfer reaction demands a proper description of the nuclear structure of the nuclear partners, represented by the ${\cal{S}}^2$, reliable optical potential for the scattering and also a detailed description of the transfer process. The usual picture of the $^{18}$O nuclei is a dineutron valence particle bound to a $^{16}$O core. Therefore, the 1NT induced by $^{18}$O occurs first by breaking the short-range pairing interaction of the two neutrons and, then, one neutron is transferred to the target nuclei. Such dynamics of pairing between two neutrons is not detailed considered into the CRC framework. The use of the \textit{psdmwkpn} interaction for the ($^{18}$O,$^{17}$O) reaction may have covered up what is, in fact, an effect of transfer mechanism instead of nuclear structure of $^{29}$Si.

%This last ingredient covers processes such as transfer followed by an excitation and the unpairing of the dineutron before the transfer, as could be the case in the ($^{18}$O,$^{17}$O) reaction.

\section{\label{conc}Conclusions}
  
The 1NT cross sections in the $^{28}$Si($^{13}$C,$^{12}$C)$^{29}$Si reaction were measured at E$_{\textnormal{lab}} = $30, and 34  MeV. Within the CRC framework, the optical potential was adjusted to describe experimental data for elastic scattering at E$_{\textnormal{lab}} = $~25, 30, and 34 MeV and the inelastic scattering at E$_{\textnormal{lab}} = $~30, and 34  MeV. The CRC calculation revealed the necessity to include a small imaginary term on the the optical potential to account for reaction channels not explicitly included in the coupling scheme. This was performed using the S\~ao Paulo potential with imaginary normalization of N$_i$~=~0.1 and indicates that some given channel has not been explicitly coupled to calculations. Elastic, inelastic and transfer data have been properly described using such configuration. The spectroscopic amplitudes obtained from the \textit{psdmod} shell model interaction provides a good description of the experimental data and in accordance with previous analysis of the (d,p) data.

\section*{Acknowledgment}
This project has received funding from CNPq, FAPERJ, FAPESP and CAPES and from INCT-FNA (Instituto Nacional de Ci\^ {e}ncia e Tecnologia- F\' isica Nuclear e Aplica\c {c}\~ {o}es). We would also like to thank the technical staff of LAFN for assisting in the maintenance and operation of the accelerator. This research has also used resources of the Laboratory of Material Analysis with Ion Beams - LAMFI-USP. The authors acknowledge the laboratory staff for assistance during the RBS experiments.

\pagebreak

%%%%\bibliography{referencias} 

\begin{thebibliography}{29}%
\makeatletter
\providecommand \@ifxundefined [1]{%
 \@ifx{#1\undefined}
}%
\providecommand \@ifnum [1]{%
 \ifnum #1\expandafter \@firstoftwo
 \else \expandafter \@secondoftwo
 \fi
}%
\providecommand \@ifx [1]{%
 \ifx #1\expandafter \@firstoftwo
 \else \expandafter \@secondoftwo
 \fi
}%
\providecommand \natexlab [1]{#1}%
\providecommand \enquote  [1]{``#1''}%
\providecommand \bibnamefont  [1]{#1}%
\providecommand \bibfnamefont [1]{#1}%
\providecommand \citenamefont [1]{#1}%
\providecommand \href@noop [0]{\@secondoftwo}%
\providecommand \href [0]{\begingroup \@sanitize@url \@href}%
\providecommand \@href[1]{\@@startlink{#1}\@@href}%
\providecommand \@@href[1]{\endgroup#1\@@endlink}%
\providecommand \@sanitize@url [0]{\catcode `\\12\catcode `\$12\catcode
  `\&12\catcode `\#12\catcode `\^12\catcode `\_12\catcode `\%12\relax}%
\providecommand \@@startlink[1]{}%
\providecommand \@@endlink[0]{}%
\providecommand \url  [0]{\begingroup\@sanitize@url \@url }%
\providecommand \@url [1]{\endgroup\@href {#1}{\urlprefix }}%
\providecommand \urlprefix  [0]{URL }%
\providecommand \Eprint [0]{\href }%
\providecommand \doibase [0]{http://dx.doi.org/}%
\providecommand \selectlanguage [0]{\@gobble}%
\providecommand \bibinfo  [0]{\@secondoftwo}%
\providecommand \bibfield  [0]{\@secondoftwo}%
\providecommand \translation [1]{[#1]}%
\providecommand \BibitemOpen [0]{}%
\providecommand \bibitemStop [0]{}%
\providecommand \bibitemNoStop [0]{.\EOS\space}%
\providecommand \EOS [0]{\spacefactor3000\relax}%
\providecommand \BibitemShut  [1]{\csname bibitem#1\endcsname}%
\let\auto@bib@innerbib\@empty
%</preamble>
\bibitem [{\citenamefont {Mermaz}\ \emph {et~al.}(1971)\citenamefont {Mermaz},
  \citenamefont {Whitten}, \citenamefont {Champlin}, \citenamefont {Howard},\
  and\ \citenamefont {Bromley}}]{MWC71}%
  \BibitemOpen
  \bibfield  {author} {\bibinfo {author} {\bibfnamefont {M.~C.}\ \bibnamefont
  {Mermaz}}, \bibinfo {author} {\bibfnamefont {C.~A.}\ \bibnamefont {Whitten}},
  \bibinfo {author} {\bibfnamefont {J.~W.}\ \bibnamefont {Champlin}}, \bibinfo
  {author} {\bibfnamefont {A.~J.}\ \bibnamefont {Howard}}, \ and\ \bibinfo
  {author} {\bibfnamefont {D.~A.}\ \bibnamefont {Bromley}},\ }\href {\doibase
  10.1103/PhysRevC.4.1778} {\bibfield  {journal} {\bibinfo  {journal} {Phys.
  Rev. C}\ }\textbf {\bibinfo {volume} {4}},\ \bibinfo {pages} {1778} (\bibinfo
  {year} {1971})}\BibitemShut {NoStop}%
\bibitem [{\citenamefont {Peterson}\ \emph {et~al.}(1983)\citenamefont
  {Peterson}, \citenamefont {Fields}, \citenamefont {Raymond}, \citenamefont
  {Thieke},\ and\ \citenamefont {Ullman}}]{PFR83}%
  \BibitemOpen
  \bibfield  {author} {\bibinfo {author} {\bibfnamefont {R.}~\bibnamefont
  {Peterson}}, \bibinfo {author} {\bibfnamefont {C.}~\bibnamefont {Fields}},
  \bibinfo {author} {\bibfnamefont {R.}~\bibnamefont {Raymond}}, \bibinfo
  {author} {\bibfnamefont {J.}~\bibnamefont {Thieke}}, \ and\ \bibinfo {author}
  {\bibfnamefont {J.}~\bibnamefont {Ullman}},\ }\href {\doibase
  https://doi.org/10.1016/0375-9474(83)90582-1} {\bibfield  {journal} {\bibinfo
   {journal} {Nuclear Physics A}\ }\textbf {\bibinfo {volume} {408}},\ \bibinfo
  {pages} {221 } (\bibinfo {year} {1983})}\BibitemShut {NoStop}%
\bibitem [{\citenamefont {Pearce}\ \emph {et~al.}(1987)\citenamefont {Pearce},
  \citenamefont {Clarke}, \citenamefont {Griffiths}, \citenamefont {Simmonds},
  \citenamefont {Barker}, \citenamefont {England}, \citenamefont {Mannion},\
  and\ \citenamefont {Ogilvie}}]{PCG87}%
  \BibitemOpen
  \bibfield  {author} {\bibinfo {author} {\bibfnamefont {K.}~\bibnamefont
  {Pearce}}, \bibinfo {author} {\bibfnamefont {N.}~\bibnamefont {Clarke}},
  \bibinfo {author} {\bibfnamefont {R.}~\bibnamefont {Griffiths}}, \bibinfo
  {author} {\bibfnamefont {P.}~\bibnamefont {Simmonds}}, \bibinfo {author}
  {\bibfnamefont {D.}~\bibnamefont {Barker}}, \bibinfo {author} {\bibfnamefont
  {J.}~\bibnamefont {England}}, \bibinfo {author} {\bibfnamefont
  {M.}~\bibnamefont {Mannion}}, \ and\ \bibinfo {author} {\bibfnamefont
  {C.}~\bibnamefont {Ogilvie}},\ }\href {\doibase
  https://doi.org/10.1016/0375-9474(87)90527-6} {\bibfield  {journal} {\bibinfo
   {journal} {Nuclear Physics A}\ }\textbf {\bibinfo {volume} {467}},\ \bibinfo
  {pages} {215 } (\bibinfo {year} {1987})}\BibitemShut {NoStop}%
\bibitem [{\citenamefont {Schumacher}\ \emph {et~al.}(1973)\citenamefont
  {Schumacher}, \citenamefont {Ueta}, \citenamefont {Duhm}, \citenamefont
  {Kubo},\ and\ \citenamefont {Klages}}]{SUD73}%
  \BibitemOpen
  \bibfield  {author} {\bibinfo {author} {\bibfnamefont {P.}~\bibnamefont
  {Schumacher}}, \bibinfo {author} {\bibfnamefont {N.}~\bibnamefont {Ueta}},
  \bibinfo {author} {\bibfnamefont {H.}~\bibnamefont {Duhm}}, \bibinfo {author}
  {\bibfnamefont {K.-I.}\ \bibnamefont {Kubo}}, \ and\ \bibinfo {author}
  {\bibfnamefont {W.}~\bibnamefont {Klages}},\ }\href {\doibase
  https://doi.org/10.1016/0375-9474(73)90824-5} {\bibfield  {journal} {\bibinfo
   {journal} {Nuclear Physics A}\ }\textbf {\bibinfo {volume} {212}},\ \bibinfo
  {pages} {573 } (\bibinfo {year} {1973})}\BibitemShut {NoStop}%
\bibitem [{\citenamefont {Hosono}\ \emph {et~al.}(1980)\citenamefont {Hosono},
  \citenamefont {Kondo}, \citenamefont {Saito}, \citenamefont {Matsuoka},
  \citenamefont {Nagamachi}, \citenamefont {Noro}, \citenamefont {Shimizu},
  \citenamefont {Kato}, \citenamefont {Okada}, \citenamefont {Ogino},\ and\
  \citenamefont {Kadota}}]{HKS80}%
  \BibitemOpen
  \bibfield  {author} {\bibinfo {author} {\bibfnamefont {K.}~\bibnamefont
  {Hosono}}, \bibinfo {author} {\bibfnamefont {M.}~\bibnamefont {Kondo}},
  \bibinfo {author} {\bibfnamefont {T.}~\bibnamefont {Saito}}, \bibinfo
  {author} {\bibfnamefont {N.}~\bibnamefont {Matsuoka}}, \bibinfo {author}
  {\bibfnamefont {S.}~\bibnamefont {Nagamachi}}, \bibinfo {author}
  {\bibfnamefont {T.}~\bibnamefont {Noro}}, \bibinfo {author} {\bibfnamefont
  {H.}~\bibnamefont {Shimizu}}, \bibinfo {author} {\bibfnamefont
  {S.}~\bibnamefont {Kato}}, \bibinfo {author} {\bibfnamefont {K.}~\bibnamefont
  {Okada}}, \bibinfo {author} {\bibfnamefont {K.}~\bibnamefont {Ogino}}, \ and\
  \bibinfo {author} {\bibfnamefont {Y.}~\bibnamefont {Kadota}},\ }\href
  {\doibase https://doi.org/10.1016/0375-9474(80)90652-1} {\bibfield  {journal}
  {\bibinfo  {journal} {Nuclear Physics A}\ }\textbf {\bibinfo {volume}
  {343}},\ \bibinfo {pages} {234 } (\bibinfo {year} {1980})}\BibitemShut
  {NoStop}%
\bibitem [{\citenamefont {Darden}\ \emph {et~al.}(1973)\citenamefont {Darden},
  \citenamefont {Sen}, \citenamefont {Hiddleston}, \citenamefont {Aymar},\ and\
  \citenamefont {Yoh}}]{DSH73}%
  \BibitemOpen
  \bibfield  {author} {\bibinfo {author} {\bibfnamefont {S.}~\bibnamefont
  {Darden}}, \bibinfo {author} {\bibfnamefont {S.}~\bibnamefont {Sen}},
  \bibinfo {author} {\bibfnamefont {H.}~\bibnamefont {Hiddleston}}, \bibinfo
  {author} {\bibfnamefont {J.}~\bibnamefont {Aymar}}, \ and\ \bibinfo {author}
  {\bibfnamefont {W.}~\bibnamefont {Yoh}},\ }\href {\doibase
  https://doi.org/10.1016/0375-9474(73)90736-7} {\bibfield  {journal} {\bibinfo
   {journal} {Nuclear Physics A}\ }\textbf {\bibinfo {volume} {208}},\ \bibinfo
  {pages} {77 } (\bibinfo {year} {1973})}\BibitemShut {NoStop}%
\bibitem [{\citenamefont {Liu}\ \emph {et~al.}(2004)\citenamefont {Liu},
  \citenamefont {Famiano}, \citenamefont {Lynch}, \citenamefont {Tsang},\ and\
  \citenamefont {Tostevin}}]{LFL04}%
  \BibitemOpen
  \bibfield  {author} {\bibinfo {author} {\bibfnamefont {X.~D.}\ \bibnamefont
  {Liu}}, \bibinfo {author} {\bibfnamefont {M.~A.}\ \bibnamefont {Famiano}},
  \bibinfo {author} {\bibfnamefont {W.~G.}\ \bibnamefont {Lynch}}, \bibinfo
  {author} {\bibfnamefont {M.~B.}\ \bibnamefont {Tsang}}, \ and\ \bibinfo
  {author} {\bibfnamefont {J.~A.}\ \bibnamefont {Tostevin}},\ }\href {\doibase
  10.1103/PhysRevC.69.064313} {\bibfield  {journal} {\bibinfo  {journal} {Phys.
  Rev. C}\ }\textbf {\bibinfo {volume} {69}},\ \bibinfo {pages} {064313}
  (\bibinfo {year} {2004})}\BibitemShut {NoStop}%
\bibitem [{\citenamefont {{Cappuzzello, F.}}\ \emph {et~al.}(2018)\citenamefont
  {{Cappuzzello, F.}}, \citenamefont {{Agodi, C.}}, \citenamefont {{Cavallaro,
  M.}}, \citenamefont {{Carbone, D.}}, \citenamefont {{Tudisco, S.}},
  \citenamefont {{Lo Presti, D.}}, \citenamefont {{Oliveira, J. R. B.}},
  \citenamefont {{Finocchiaro, P.}}, \citenamefont {{Colonna, M.}},
  \citenamefont {{Rifuggiato, D.}}, \citenamefont {{Calabretta, L.}},
  \citenamefont {{Calvo, D.}}, \citenamefont {{Pandola, L.}}, \citenamefont
  {{Acosta, L.}}, \citenamefont {{Auerbach, N.}}, \citenamefont {{Bellone,
  J.}}, \citenamefont {{Bijker, R.}}, \citenamefont {{Bonanno, D.}},
  \citenamefont {{Bongiovanni, D.}}, \citenamefont {{Borello-Lewin, T.}},
  \citenamefont {{Boztosun, I.}}, \citenamefont {{Brunasso, O.}}, \citenamefont
  {{Burrello, S.}}, \citenamefont {{Calabrese, S.}}, \citenamefont {{Calanna,
  A.}}, \citenamefont {{Ch\'avez Lomel\'{\i}, E. R.}}, \citenamefont
  {{D\'{}Agostino, G.}}, \citenamefont {{De Faria, P. N.}}, \citenamefont {{De
  Geronimo, G.}}, \citenamefont {{Delaunay, F.}}, \citenamefont {{Deshmukh,
  N.}}, \citenamefont {{Ferreira, J. L.}}, \citenamefont {{Fisichella, M.}},
  \citenamefont {{Foti, A.}}, \citenamefont {{Gallo, G.}}, \citenamefont
  {{Garcia-Tecocoatzi, H.}}, \citenamefont {{Greco, V.}}, \citenamefont
  {{Hacisalihoglu, A.}}, \citenamefont {{Iazzi, F.}}, \citenamefont {{Introzzi,
  R.}}, \citenamefont {{Lanzalone, G.}}, \citenamefont {{Lay, J. A.}},
  \citenamefont {{La Via, F.}}, \citenamefont {{Lenske, H.}}, \citenamefont
  {{Linares, R.}}, \citenamefont {{Litrico, G.}}, \citenamefont {{Longhitano,
  F.}}, \citenamefont {{Lubian, J.}}, \citenamefont {{Medina, N. H.}},
  \citenamefont {{Mendes, D. R.}}, \citenamefont {{Moralles, M.}},
  \citenamefont {{Muoio, A.}}, \citenamefont {{Pakou, A.}}, \citenamefont
  {{Petrascu, H.}}, \citenamefont {{Pinna, F.}}, \citenamefont {{Reito, S.}},
  \citenamefont {{Russo, A. D.}}, \citenamefont {{Russo, G.}}, \citenamefont
  {{Santagati, G.}}, \citenamefont {{Santopinto, E.}}, \citenamefont {{Santos,
  R. B. B.}}, \citenamefont {{Sgouros, O.}}, \citenamefont {{da Silveira, M. A.
  G.}}, \citenamefont {{Solakci, S. O.}}, \citenamefont {{Souliotis, G.}},
  \citenamefont {{Soukeras, V.}}, \citenamefont {{Spatafora, A.}},
  \citenamefont {{Torresi, D.}}, \citenamefont {{Magana Vsevolodovna, R.}},
  \citenamefont {{Yildirim, A.}},\ and\ \citenamefont {{Zagatto, V. A.
  B.}}}]{CAC18}%
  \BibitemOpen
  \bibfield  {author} {\bibinfo {author} {\bibnamefont {{Cappuzzello, F.}}},
  \bibinfo {author} {\bibnamefont {{Agodi, C.}}}, \bibinfo {author}
  {\bibnamefont {{Cavallaro, M.}}}, \bibinfo {author} {\bibnamefont {{Carbone,
  D.}}}, \bibinfo {author} {\bibnamefont {{Tudisco, S.}}}, \bibinfo {author}
  {\bibnamefont {{Lo Presti, D.}}}, \bibinfo {author} {\bibnamefont {{Oliveira,
  J. R. B.}}}, \bibinfo {author} {\bibnamefont {{Finocchiaro, P.}}}, \bibinfo
  {author} {\bibnamefont {{Colonna, M.}}}, \bibinfo {author} {\bibnamefont
  {{Rifuggiato, D.}}}, \bibinfo {author} {\bibnamefont {{Calabretta, L.}}},
  \bibinfo {author} {\bibnamefont {{Calvo, D.}}}, \bibinfo {author}
  {\bibnamefont {{Pandola, L.}}}, \bibinfo {author} {\bibnamefont {{Acosta,
  L.}}}, \bibinfo {author} {\bibnamefont {{Auerbach, N.}}}, \bibinfo {author}
  {\bibnamefont {{Bellone, J.}}}, \bibinfo {author} {\bibnamefont {{Bijker,
  R.}}}, \bibinfo {author} {\bibnamefont {{Bonanno, D.}}}, \bibinfo {author}
  {\bibnamefont {{Bongiovanni, D.}}}, \bibinfo {author} {\bibnamefont
  {{Borello-Lewin, T.}}}, \bibinfo {author} {\bibnamefont {{Boztosun, I.}}},
  \bibinfo {author} {\bibnamefont {{Brunasso, O.}}}, \bibinfo {author}
  {\bibnamefont {{Burrello, S.}}}, \bibinfo {author} {\bibnamefont {{Calabrese,
  S.}}}, \bibinfo {author} {\bibnamefont {{Calanna, A.}}}, \bibinfo {author}
  {\bibnamefont {{Ch\'avez Lomel\'{\i}, E. R.}}}, \bibinfo {author}
  {\bibnamefont {{D\'{}Agostino, G.}}}, \bibinfo {author} {\bibnamefont {{De
  Faria, P. N.}}}, \bibinfo {author} {\bibnamefont {{De Geronimo, G.}}},
  \bibinfo {author} {\bibnamefont {{Delaunay, F.}}}, \bibinfo {author}
  {\bibnamefont {{Deshmukh, N.}}}, \bibinfo {author} {\bibnamefont {{Ferreira,
  J. L.}}}, \bibinfo {author} {\bibnamefont {{Fisichella, M.}}}, \bibinfo
  {author} {\bibnamefont {{Foti, A.}}}, \bibinfo {author} {\bibnamefont
  {{Gallo, G.}}}, \bibinfo {author} {\bibnamefont {{Garcia-Tecocoatzi, H.}}},
  \bibinfo {author} {\bibnamefont {{Greco, V.}}}, \bibinfo {author}
  {\bibnamefont {{Hacisalihoglu, A.}}}, \bibinfo {author} {\bibnamefont
  {{Iazzi, F.}}}, \bibinfo {author} {\bibnamefont {{Introzzi, R.}}}, \bibinfo
  {author} {\bibnamefont {{Lanzalone, G.}}}, \bibinfo {author} {\bibnamefont
  {{Lay, J. A.}}}, \bibinfo {author} {\bibnamefont {{La Via, F.}}}, \bibinfo
  {author} {\bibnamefont {{Lenske, H.}}}, \bibinfo {author} {\bibnamefont
  {{Linares, R.}}}, \bibinfo {author} {\bibnamefont {{Litrico, G.}}}, \bibinfo
  {author} {\bibnamefont {{Longhitano, F.}}}, \bibinfo {author} {\bibnamefont
  {{Lubian, J.}}}, \bibinfo {author} {\bibnamefont {{Medina, N. H.}}}, \bibinfo
  {author} {\bibnamefont {{Mendes, D. R.}}}, \bibinfo {author} {\bibnamefont
  {{Moralles, M.}}}, \bibinfo {author} {\bibnamefont {{Muoio, A.}}}, \bibinfo
  {author} {\bibnamefont {{Pakou, A.}}}, \bibinfo {author} {\bibnamefont
  {{Petrascu, H.}}}, \bibinfo {author} {\bibnamefont {{Pinna, F.}}}, \bibinfo
  {author} {\bibnamefont {{Reito, S.}}}, \bibinfo {author} {\bibnamefont
  {{Russo, A. D.}}}, \bibinfo {author} {\bibnamefont {{Russo, G.}}}, \bibinfo
  {author} {\bibnamefont {{Santagati, G.}}}, \bibinfo {author} {\bibnamefont
  {{Santopinto, E.}}}, \bibinfo {author} {\bibnamefont {{Santos, R. B. B.}}},
  \bibinfo {author} {\bibnamefont {{Sgouros, O.}}}, \bibinfo {author}
  {\bibnamefont {{da Silveira, M. A. G.}}}, \bibinfo {author} {\bibnamefont
  {{Solakci, S. O.}}}, \bibinfo {author} {\bibnamefont {{Souliotis, G.}}},
  \bibinfo {author} {\bibnamefont {{Soukeras, V.}}}, \bibinfo {author}
  {\bibnamefont {{Spatafora, A.}}}, \bibinfo {author} {\bibnamefont {{Torresi,
  D.}}}, \bibinfo {author} {\bibnamefont {{Magana Vsevolodovna, R.}}}, \bibinfo
  {author} {\bibnamefont {{Yildirim, A.}}}, \ and\ \bibinfo {author}
  {\bibnamefont {{Zagatto, V. A. B.}}},\ }\href {\doibase
  10.1140/epja/i2018-12509-3} {\bibfield  {journal} {\bibinfo  {journal} {Eur.
  Phys. J. A}\ }\textbf {\bibinfo {volume} {54}},\ \bibinfo {pages} {72}
  (\bibinfo {year} {2018})}\BibitemShut {NoStop}%
\bibitem [{\citenamefont {Bailey}\ \emph {et~al.}(2016)\citenamefont {Bailey},
  \citenamefont {Timofeyuk},\ and\ \citenamefont {Tostevin}}]{BTT16}%
  \BibitemOpen
  \bibfield  {author} {\bibinfo {author} {\bibfnamefont {G.~W.}\ \bibnamefont
  {Bailey}}, \bibinfo {author} {\bibfnamefont {N.~K.}\ \bibnamefont
  {Timofeyuk}}, \ and\ \bibinfo {author} {\bibfnamefont {J.~A.}\ \bibnamefont
  {Tostevin}},\ }\href {\doibase 10.1103/PhysRevLett.117.162502} {\bibfield
  {journal} {\bibinfo  {journal} {Phys. Rev. Lett.}\ }\textbf {\bibinfo
  {volume} {117}},\ \bibinfo {pages} {162502} (\bibinfo {year}
  {2016})}\BibitemShut {NoStop}%
\bibitem [{\citenamefont {Timofeyuk}\ and\ \citenamefont
  {Johnson}(2013)}]{TiJ13}%
  \BibitemOpen
  \bibfield  {author} {\bibinfo {author} {\bibfnamefont {N.~K.}\ \bibnamefont
  {Timofeyuk}}\ and\ \bibinfo {author} {\bibfnamefont {R.~C.}\ \bibnamefont
  {Johnson}},\ }\href {\doibase 10.1103/PhysRevLett.110.112501} {\bibfield
  {journal} {\bibinfo  {journal} {Phys. Rev. Lett.}\ }\textbf {\bibinfo
  {volume} {110}},\ \bibinfo {pages} {112501} (\bibinfo {year}
  {2013})}\BibitemShut {NoStop}%
\bibitem [{\citenamefont {Ross}\ \emph {et~al.}(2016)\citenamefont {Ross},
  \citenamefont {Titus},\ and\ \citenamefont {Nunes}}]{RTN16}%
  \BibitemOpen
  \bibfield  {author} {\bibinfo {author} {\bibfnamefont {A.}~\bibnamefont
  {Ross}}, \bibinfo {author} {\bibfnamefont {L.~J.}\ \bibnamefont {Titus}}, \
  and\ \bibinfo {author} {\bibfnamefont {F.~M.}\ \bibnamefont {Nunes}},\ }\href
  {\doibase 10.1103/PhysRevC.94.014607} {\bibfield  {journal} {\bibinfo
  {journal} {Phys. Rev. C}\ }\textbf {\bibinfo {volume} {94}},\ \bibinfo
  {pages} {014607} (\bibinfo {year} {2016})}\BibitemShut {NoStop}%
\bibitem [{\citenamefont {Linares}\ \emph {et~al.}(2018)\citenamefont
  {Linares}, \citenamefont {Ermamatov}, \citenamefont {Lubian}, \citenamefont
  {Cappuzzello}, \citenamefont {Carbone}, \citenamefont {Cardozo},
  \citenamefont {Cavallaro}, \citenamefont {Ferreira}, \citenamefont {Foti},
  \citenamefont {Gargano}, \citenamefont {Paes}, \citenamefont {Santagati},\
  and\ \citenamefont {Zagatto}}]{LEL18}%
  \BibitemOpen
  \bibfield  {author} {\bibinfo {author} {\bibfnamefont {R.}~\bibnamefont
  {Linares}}, \bibinfo {author} {\bibfnamefont {M.~J.}\ \bibnamefont
  {Ermamatov}}, \bibinfo {author} {\bibfnamefont {J.}~\bibnamefont {Lubian}},
  \bibinfo {author} {\bibfnamefont {F.}~\bibnamefont {Cappuzzello}}, \bibinfo
  {author} {\bibfnamefont {D.}~\bibnamefont {Carbone}}, \bibinfo {author}
  {\bibfnamefont {E.~N.}\ \bibnamefont {Cardozo}}, \bibinfo {author}
  {\bibfnamefont {M.}~\bibnamefont {Cavallaro}}, \bibinfo {author}
  {\bibfnamefont {J.~L.}\ \bibnamefont {Ferreira}}, \bibinfo {author}
  {\bibfnamefont {A.}~\bibnamefont {Foti}}, \bibinfo {author} {\bibfnamefont
  {A.}~\bibnamefont {Gargano}}, \bibinfo {author} {\bibfnamefont
  {B.}~\bibnamefont {Paes}}, \bibinfo {author} {\bibfnamefont {G.}~\bibnamefont
  {Santagati}}, \ and\ \bibinfo {author} {\bibfnamefont {V.~A.~B.}\
  \bibnamefont {Zagatto}},\ }\href {\doibase 10.1103/PhysRevC.98.054615}
  {\bibfield  {journal} {\bibinfo  {journal} {Phys. Rev. C}\ }\textbf {\bibinfo
  {volume} {98}},\ \bibinfo {pages} {054615} (\bibinfo {year}
  {2018})}\BibitemShut {NoStop}%
\bibitem [{\citenamefont {Cardozo}\ \emph {et~al.}(2018)\citenamefont
  {Cardozo}, \citenamefont {Lubian}, \citenamefont {Linares}, \citenamefont
  {Cappuzzello}, \citenamefont {Carbone}, \citenamefont {Cavallaro},
  \citenamefont {Ferreira}, \citenamefont {Gargano}, \citenamefont {Paes},\
  and\ \citenamefont {Santagati}}]{CLL18}%
  \BibitemOpen
  \bibfield  {author} {\bibinfo {author} {\bibfnamefont {E.~N.}\ \bibnamefont
  {Cardozo}}, \bibinfo {author} {\bibfnamefont {J.}~\bibnamefont {Lubian}},
  \bibinfo {author} {\bibfnamefont {R.}~\bibnamefont {Linares}}, \bibinfo
  {author} {\bibfnamefont {F.}~\bibnamefont {Cappuzzello}}, \bibinfo {author}
  {\bibfnamefont {D.}~\bibnamefont {Carbone}}, \bibinfo {author} {\bibfnamefont
  {M.}~\bibnamefont {Cavallaro}}, \bibinfo {author} {\bibfnamefont {J.~L.}\
  \bibnamefont {Ferreira}}, \bibinfo {author} {\bibfnamefont {A.}~\bibnamefont
  {Gargano}}, \bibinfo {author} {\bibfnamefont {B.}~\bibnamefont {Paes}}, \
  and\ \bibinfo {author} {\bibfnamefont {G.}~\bibnamefont {Santagati}},\ }\href
  {\doibase 10.1103/PhysRevC.97.064611} {\bibfield  {journal} {\bibinfo
  {journal} {Phys. Rev. C}\ }\textbf {\bibinfo {volume} {97}},\ \bibinfo
  {pages} {064611} (\bibinfo {year} {2018})}\BibitemShut {NoStop}%
\bibitem [{\citenamefont {Gasques}\ \emph {et~al.}(2018)\citenamefont
  {Gasques}, \citenamefont {Freitas}, \citenamefont {Chamon}, \citenamefont
  {Oliveira}, \citenamefont {Medina}, \citenamefont {Scarduelli}, \citenamefont
  {Rossi}, \citenamefont {Alvarez}, \citenamefont {Zagatto}, \citenamefont
  {Lubian}, \citenamefont {Nobre}, \citenamefont {Padron},\ and\ \citenamefont
  {Carlson}}]{GFC18}%
  \BibitemOpen
  \bibfield  {author} {\bibinfo {author} {\bibfnamefont {L.~R.}\ \bibnamefont
  {Gasques}}, \bibinfo {author} {\bibfnamefont {A.~S.}\ \bibnamefont
  {Freitas}}, \bibinfo {author} {\bibfnamefont {L.~C.}\ \bibnamefont {Chamon}},
  \bibinfo {author} {\bibfnamefont {J.~R.~B.}\ \bibnamefont {Oliveira}},
  \bibinfo {author} {\bibfnamefont {N.~H.}\ \bibnamefont {Medina}}, \bibinfo
  {author} {\bibfnamefont {V.}~\bibnamefont {Scarduelli}}, \bibinfo {author}
  {\bibfnamefont {E.~S.}\ \bibnamefont {Rossi}}, \bibinfo {author}
  {\bibfnamefont {M.~A.~G.}\ \bibnamefont {Alvarez}}, \bibinfo {author}
  {\bibfnamefont {V.~A.~B.}\ \bibnamefont {Zagatto}}, \bibinfo {author}
  {\bibfnamefont {J.}~\bibnamefont {Lubian}}, \bibinfo {author} {\bibfnamefont
  {G.~P.~A.}\ \bibnamefont {Nobre}}, \bibinfo {author} {\bibfnamefont
  {I.}~\bibnamefont {Padron}}, \ and\ \bibinfo {author} {\bibfnamefont {B.~V.}\
  \bibnamefont {Carlson}},\ }\href {\doibase 10.1103/PhysRevC.97.034629}
  {\bibfield  {journal} {\bibinfo  {journal} {Phys. Rev. C}\ }\textbf {\bibinfo
  {volume} {97}},\ \bibinfo {pages} {034629} (\bibinfo {year}
  {2018})}\BibitemShut {NoStop}%
\bibitem [{\citenamefont {Tarasov}\ and\ \citenamefont {Bazin}(2008)}]{TaB08}%
  \BibitemOpen
  \bibfield  {author} {\bibinfo {author} {\bibfnamefont {O.}~\bibnamefont
  {Tarasov}}\ and\ \bibinfo {author} {\bibfnamefont {D.}~\bibnamefont
  {Bazin}},\ }\href {\doibase https://doi.org/10.1016/j.nimb.2008.05.110}
  {\bibfield  {journal} {\bibinfo  {journal} {Nucl. Instrum. Methods Phys. Res.
  Sect. B}\ }\textbf {\bibinfo {volume} {266}},\ \bibinfo {pages} {4657 }
  (\bibinfo {year} {2008})},\ \bibinfo {note} {proceedings of the XVth
  International Conference on Electromagnetic Isotope Separators and Techniques
  Related to their Applications}\BibitemShut {NoStop}%
\bibitem [{\citenamefont {Gavron}(1980)}]{Gav80}%
  \BibitemOpen
  \bibfield  {author} {\bibinfo {author} {\bibfnamefont {A.}~\bibnamefont
  {Gavron}},\ }\href {\doibase 10.1103/PhysRevC.21.230} {\bibfield  {journal}
  {\bibinfo  {journal} {Phys. Rev. C}\ }\textbf {\bibinfo {volume} {21}},\
  \bibinfo {pages} {230} (\bibinfo {year} {1980})}\BibitemShut {NoStop}%
\bibitem [{\citenamefont {Thompson}(1988)}]{Tho88}%
  \BibitemOpen
  \bibfield  {author} {\bibinfo {author} {\bibfnamefont {I.~J.}\ \bibnamefont
  {Thompson}},\ }\href@noop {} {\bibfield  {journal} {\bibinfo  {journal}
  {Comput. Phys. Rep.}\ }\textbf {\bibinfo {volume} {7}},\ \bibinfo {pages}
  {167} (\bibinfo {year} {1988})}\BibitemShut {NoStop}%
\bibitem [{\citenamefont {Raman}\ \emph {et~al.}(2001)\citenamefont {Raman},
  \citenamefont {Nestor},\ and\ \citenamefont {Tikkanen}}]{RNT01}%
  \BibitemOpen
  \bibfield  {author} {\bibinfo {author} {\bibfnamefont {S.}~\bibnamefont
  {Raman}}, \bibinfo {author} {\bibfnamefont {C.}~\bibnamefont {Nestor},
  \bibfnamefont {Jr.}}, \ and\ \bibinfo {author} {\bibfnamefont
  {P.}~\bibnamefont {Tikkanen}},\ }\href@noop {} {\bibfield  {journal}
  {\bibinfo  {journal} {Atomic Data and Nuclear Data Tables}\ }\textbf
  {\bibinfo {volume} {78}},\ \bibinfo {pages} {1} (\bibinfo {year}
  {2001})}\BibitemShut {NoStop}%
\bibitem [{\citenamefont {Ajzenberg-Selove}(1991)}]{Ajz91}%
  \BibitemOpen
  \bibfield  {author} {\bibinfo {author} {\bibfnamefont {F.}~\bibnamefont
  {Ajzenberg-Selove}},\ }\href {\doibase
  https://doi.org/10.1016/0375-9474(91)90446-D} {\bibfield  {journal} {\bibinfo
   {journal} {Nuclear Physics A}\ }\textbf {\bibinfo {volume} {523}},\ \bibinfo
  {pages} {1 } (\bibinfo {year} {1991})}\BibitemShut {NoStop}%
\bibitem [{\citenamefont {Cavallaro}\ \emph {et~al.}(2013)\citenamefont
  {Cavallaro}, \citenamefont {Cappuzzello}, \citenamefont {Bond\`{i}},
  \citenamefont {Carbone}, \citenamefont {Garcia}, \citenamefont {Gargano},
  \citenamefont {Lenzi}, \citenamefont {Lubian}, \citenamefont {Agodi},
  \citenamefont {Azaiez}, \citenamefont {De~Napoli}, \citenamefont {Foti},
  \citenamefont {Franchoo}, \citenamefont {Linares}, \citenamefont {Nicolosi},
  \citenamefont {Niikura}, \citenamefont {Scarpaci},\ and\ \citenamefont
  {Tropea}}]{CCB13}%
  \BibitemOpen
  \bibfield  {author} {\bibinfo {author} {\bibfnamefont {M.}~\bibnamefont
  {Cavallaro}}, \bibinfo {author} {\bibfnamefont {F.}~\bibnamefont
  {Cappuzzello}}, \bibinfo {author} {\bibfnamefont {M.}~\bibnamefont
  {Bond\`{i}}}, \bibinfo {author} {\bibfnamefont {D.}~\bibnamefont {Carbone}},
  \bibinfo {author} {\bibfnamefont {V.~N.}\ \bibnamefont {Garcia}}, \bibinfo
  {author} {\bibfnamefont {A.}~\bibnamefont {Gargano}}, \bibinfo {author}
  {\bibfnamefont {S.~M.}\ \bibnamefont {Lenzi}}, \bibinfo {author}
  {\bibfnamefont {J.}~\bibnamefont {Lubian}}, \bibinfo {author} {\bibfnamefont
  {C.}~\bibnamefont {Agodi}}, \bibinfo {author} {\bibfnamefont
  {F.}~\bibnamefont {Azaiez}}, \bibinfo {author} {\bibfnamefont
  {M.}~\bibnamefont {De~Napoli}}, \bibinfo {author} {\bibfnamefont
  {A.}~\bibnamefont {Foti}}, \bibinfo {author} {\bibfnamefont {S.}~\bibnamefont
  {Franchoo}}, \bibinfo {author} {\bibfnamefont {R.}~\bibnamefont {Linares}},
  \bibinfo {author} {\bibfnamefont {D.}~\bibnamefont {Nicolosi}}, \bibinfo
  {author} {\bibfnamefont {M.}~\bibnamefont {Niikura}}, \bibinfo {author}
  {\bibfnamefont {J.~A.}\ \bibnamefont {Scarpaci}}, \ and\ \bibinfo {author}
  {\bibfnamefont {S.}~\bibnamefont {Tropea}},\ }\href@noop {} {\bibfield
  {journal} {\bibinfo  {journal} {Phys. Rev. C}\ }\textbf {\bibinfo {volume}
  {88}},\ \bibinfo {pages} {054601} (\bibinfo {year} {2013})}\BibitemShut
  {NoStop}%
\bibitem [{\citenamefont {Rae}(2008)}]{Rae08}%
  \BibitemOpen
  \bibfield  {author} {\bibinfo {author} {\bibfnamefont {W.~D.~M.}\
  \bibnamefont {Rae}},\ }\href {http://www.garsington.eclipse.co.uk} {\bibfield
   {journal} {\bibinfo  {journal} {http://www.garsington.eclipse.co.uk}\ }
  (\bibinfo {year} {2008})}\BibitemShut {NoStop}%
\bibitem [{\citenamefont {Utsuno}\ and\ \citenamefont {Chiba}(2011)}]{UC11}%
  \BibitemOpen
  \bibfield  {author} {\bibinfo {author} {\bibfnamefont {Y.}~\bibnamefont
  {Utsuno}}\ and\ \bibinfo {author} {\bibfnamefont {S.}~\bibnamefont {Chiba}},\
  }\href@noop {} {\bibfield  {journal} {\bibinfo  {journal} {Phys. Rev. C}\
  }\textbf {\bibinfo {volume} {83}},\ \bibinfo {pages} {021301(R)} (\bibinfo
  {year} {2011})}\BibitemShut {NoStop}%
\bibitem [{\citenamefont {Taqi Al-Bayati}\ and\ \citenamefont
  {Darwesh}(2013)}]{BaD13}%
  \BibitemOpen
  \bibfield  {author} {\bibinfo {author} {\bibfnamefont {A.~H.}\ \bibnamefont
  {Taqi Al-Bayati}}\ and\ \bibinfo {author} {\bibfnamefont {S.~S.}\
  \bibnamefont {Darwesh}},\ }\href {\doibase 10.1063/1.4849222} {\bibfield
  {journal} {\bibinfo  {journal} {AIP Conference Proceedings}\ }\textbf
  {\bibinfo {volume} {1569}},\ \bibinfo {pages} {27} (\bibinfo {year}
  {2013})},\ \Eprint
  {http://arxiv.org/abs/https://aip.scitation.org/doi/pdf/10.1063/1.4849222}
  {https://aip.scitation.org/doi/pdf/10.1063/1.4849222} \BibitemShut {NoStop}%
\bibitem [{\citenamefont {Cohen}\ and\ \citenamefont {Kurath}(1965)}]{CoK65}%
  \BibitemOpen
  \bibfield  {author} {\bibinfo {author} {\bibfnamefont {S.}~\bibnamefont
  {Cohen}}\ and\ \bibinfo {author} {\bibfnamefont {D.}~\bibnamefont {Kurath}},\
  }\href {\doibase https://doi.org/10.1016/0029-5582(65)90148-3} {\bibfield
  {journal} {\bibinfo  {journal} {Nuclear Physics}\ }\textbf {\bibinfo {volume}
  {73}},\ \bibinfo {pages} {1 } (\bibinfo {year} {1965})}\BibitemShut {NoStop}%
\bibitem [{\citenamefont {Wildenthal}(1984)}]{Wil84}%
  \BibitemOpen
  \bibfield  {author} {\bibinfo {author} {\bibfnamefont {B.}~\bibnamefont
  {Wildenthal}},\ }\href {\doibase
  https://doi.org/10.1016/0146-6410(84)90011-5} {\bibfield  {journal} {\bibinfo
   {journal} {Progress in Particle and Nuclear Physics}\ }\textbf {\bibinfo
  {volume} {11}},\ \bibinfo {pages} {5 } (\bibinfo {year} {1984})}\BibitemShut
  {NoStop}%
\bibitem [{\citenamefont {Millener}\ and\ \citenamefont
  {Kurath}(1975)}]{MiK75}%
  \BibitemOpen
  \bibfield  {author} {\bibinfo {author} {\bibfnamefont {D.}~\bibnamefont
  {Millener}}\ and\ \bibinfo {author} {\bibfnamefont {D.}~\bibnamefont
  {Kurath}},\ }\href {\doibase https://doi.org/10.1016/0375-9474(75)90683-1}
  {\bibfield  {journal} {\bibinfo  {journal} {Nuclear Physics A}\ }\textbf
  {\bibinfo {volume} {255}},\ \bibinfo {pages} {315 } (\bibinfo {year}
  {1975})}\BibitemShut {NoStop}%
\bibitem [{\citenamefont {Chamon}\ \emph {et~al.}(2002)\citenamefont {Chamon},
  \citenamefont {Carlson}, \citenamefont {Gasques}, \citenamefont {Pereira},
  \citenamefont {De~Conti}, \citenamefont {Alvarez}, \citenamefont {Hussein},
  \citenamefont {C\^andido~Ribeiro}, \citenamefont {Rossi~Jr.},\ and\
  \citenamefont {Silva}}]{CCG02}%
  \BibitemOpen
  \bibfield  {author} {\bibinfo {author} {\bibfnamefont {L.~C.}\ \bibnamefont
  {Chamon}}, \bibinfo {author} {\bibfnamefont {B.~V.}\ \bibnamefont {Carlson}},
  \bibinfo {author} {\bibfnamefont {L.~R.}\ \bibnamefont {Gasques}}, \bibinfo
  {author} {\bibfnamefont {D.}~\bibnamefont {Pereira}}, \bibinfo {author}
  {\bibfnamefont {C.}~\bibnamefont {De~Conti}}, \bibinfo {author}
  {\bibfnamefont {M.~A.~G.}\ \bibnamefont {Alvarez}}, \bibinfo {author}
  {\bibfnamefont {M.~S.}\ \bibnamefont {Hussein}}, \bibinfo {author}
  {\bibfnamefont {M.~A.}\ \bibnamefont {C\^andido~Ribeiro}}, \bibinfo {author}
  {\bibfnamefont {E.~S.}\ \bibnamefont {Rossi~Jr.}}, \ and\ \bibinfo {author}
  {\bibfnamefont {C.~P.}\ \bibnamefont {Silva}},\ }\href@noop {} {\bibfield
  {journal} {\bibinfo  {journal} {Phys. Rev. C}\ }\textbf {\bibinfo {volume}
  {66}},\ \bibinfo {pages} {014610} (\bibinfo {year} {2002})}\BibitemShut
  {NoStop}%
\bibitem [{\citenamefont {Alvarez}\ \emph {et~al.}(2003)\citenamefont
  {Alvarez}, \citenamefont {Chamon}, \citenamefont {Hussein}, \citenamefont
  {Pereira}, \citenamefont {Gasques}, \citenamefont {Rossi},\ and\
  \citenamefont {Silva}}]{ACH03}%
  \BibitemOpen
  \bibfield  {author} {\bibinfo {author} {\bibfnamefont {M.~A.~G.}\
  \bibnamefont {Alvarez}}, \bibinfo {author} {\bibfnamefont {L.~C.}\
  \bibnamefont {Chamon}}, \bibinfo {author} {\bibfnamefont {M.~S.}\
  \bibnamefont {Hussein}}, \bibinfo {author} {\bibfnamefont {D.}~\bibnamefont
  {Pereira}}, \bibinfo {author} {\bibfnamefont {L.~R.}\ \bibnamefont
  {Gasques}}, \bibinfo {author} {\bibfnamefont {E.~S.}\ \bibnamefont {Rossi}},
  \ and\ \bibinfo {author} {\bibfnamefont {C.~P.}\ \bibnamefont {Silva}},\
  }\href@noop {} {\bibfield  {journal} {\bibinfo  {journal} {Nucl. Phys. A}\
  }\textbf {\bibinfo {volume} {723}},\ \bibinfo {pages} {93} (\bibinfo {year}
  {2003})}\BibitemShut {NoStop}%
\bibitem [{\citenamefont {Tsang}\ \emph {et~al.}(2005)\citenamefont {Tsang},
  \citenamefont {Lee},\ and\ \citenamefont {Lynch}}]{TLL05}%
  \BibitemOpen
  \bibfield  {author} {\bibinfo {author} {\bibfnamefont {M.~B.}\ \bibnamefont
  {Tsang}}, \bibinfo {author} {\bibfnamefont {J.}~\bibnamefont {Lee}}, \ and\
  \bibinfo {author} {\bibfnamefont {W.~G.}\ \bibnamefont {Lynch}},\ }\href
  {\doibase 10.1103/PhysRevLett.95.222501} {\bibfield  {journal} {\bibinfo
  {journal} {Phys. Rev. Lett.}\ }\textbf {\bibinfo {volume} {95}},\ \bibinfo
  {pages} {222501} (\bibinfo {year} {2005})}\BibitemShut {NoStop}%
\end{thebibliography}

%

%%%%%%%%%%%%%%%%%%%%%%%%%%%%%%%%%%%%%%%%%%%%%%%%%%%%%%%%%%%%%%%%%%%%%%%%%%%%%%%%%%%%%%

\end{document}